\documentclass[%
 reprint,
 amsmath,amssymb,
 aps,
]{revtex4-1}

\usepackage{graphicx}
\usepackage{dcolumn}
\usepackage{bm}

\begin{document}

\preprint{APS/123-QED}

\title{Structure of inert layer $^{4}$He adsorbed on a mesoporous silica}

\author{Junko Taniguchi}
 \email{tany@phys.uec.ac.jp}
\author{Kizashi Mikami, and Masaru Suzuki}
\affiliation{%
 University of Electro-Communications, Chofu, Tokyo, Japan
}%




\date{\today}

\begin{abstract}

We have studied the structure of inert layer $^{4}$He adsorbed on a mesoporous silica (FSM-16), 
by the vapor pressure and heat capacity measurements. 
The heat capacity shows a Schottky-like peak due to the excitation of a part of localized solid to fluid. 
We analyzed the heat capacity over a wide temperature region based on the model including the contribution 
of the localized solid and excited fluid and clarified that the excited fluid coexists with the localized solid 
at high temperature. As the areal density approaches that at which superfluid appears ($n_{\mathrm{C}}$), 
the fluid amount is likely to go to zero, suggesting a possibility that the inert layer is solidified just below $n_{\mathrm{C}}$.

\end{abstract}

\pacs{Valid PACS appear here}
\maketitle


\section{Introduction}

$^{4}$He films adsorbed on various substrates have been extensively studied, since it can be a 
stage of novel superfluidity. In fact, two-dimensional Kosterlitz-Thouless transition\cite{Bishop_1978}, 
its size effect\cite{Shirahama_1990}, dimensional crossover in superfluidity\cite{Matsushita_2017}, 
superfluidity intertwined with a density wave order\cite{Nyeki_2017} 
were reported. On the other hand, the film structure itself has also attracted researchers' interests. 
On graphite, which is known by homogeneous adsorption potential, layer-by-layer growth of films 
were confirmed up to 6-atomic layers\cite{Zimmerli_1992}, and very complicated phase diagram for 
the first and the second layer was clarified\cite{Shick_1980,Greywall}. 
On other substrates, such complicated phase diagram was not reported, due to 
the heterogeneous adsorption potential compared with graphite. On these substrates, it is thought 
that an inert layer is formed below the areal density at which a superfluid layer appears ($n_{\mathrm{c}}$). 

Among various heterogeneous substrates, on a vycor glass, the film structure has been studied 
in detail based on heat capacity measurements\cite{Tait_1979}. 
Tait and Reppy explained the observed bend in heat capacity by the excitation 
of a part of $^{4}$He from localized solid islands to delocalized gas area. They insisted that, 
on the high temperature side of the bend, the excited gas coexists with solid islands, 
and covers only a fraction of the total surface area, due to a lateral pressure caused by the 
distribution of long-range adsorption potential. 

Recently, Toda et al. studied the film structure of $^{4}$He adsorbed on a mesoporous silica 
called HMM-2 based on vapor pressure and heat capacity measurements\cite{Toda_2009}. 
They observed the similar heat capacity for the inert layer. 
They insisted that above the areal density of the first layer completion, the heat capacity on 
the high-temperature side of the bend can be explained as a 
normal fluid which shows an amorphous-solid like temperature dependence\cite{Andreev_1978}. 
To confirm this, they first qualitatively evaluated the density state of two-level systems (TLS) in 
the amorphous solid state from the vapor pressure data. 

In the previous work, the structure of the inert layer was discussed based on the heat capacity 
of limited temperature or areal density regions. 
Our motivation is to understand its structure more comprehensively and quantitatively. 
Thus, we chose the mesoporous silica called FSM-16, whose distribution of adsorption 
potential is theoretically studied\cite{Rossi_2006}, and performed the heat capacity and the vapor 
pressure measurements. We analyzed the heat capacity of the inert layer 
in a wide temperature region (0.18-4.5~K), based on the model including the contribution of the 
localized solid and excited fluid. 
The results support the picture that the excited fluid coexists with the localized solid at high temperature. 
On the other hand, its amount is likely to go to zero as the areal density approaches $n_{\mathrm{C}}$, suggesting a new possibility that the inert layer is solidified just below $n_{\mathrm{C}}$. 

\section{Experiments}

The mesoporous silica called FSM was first synthesized by Inagaki {\it et al}. in Toyota 
Central R\&D Labs., Inc. Japan.\cite{Inagaki_1993} It forms a honeycomb 
structure of a 1D uniform nanometer-size straight channel without 
interconnection. Using an organic molecule as a template, the diameter 
of the channel was precisely controlled. Its homogeneity was confirmed 
by the transmission electron micrograph and the X-ray diffraction (XRD) 
pattern.\cite{Inagaki_1996} The sample we used is the one with 2.8-nm channel. 

For the heat capacity measurements, we used the cell which was used in the previous 
heat capacity measurements for pressurized liquid $^{4}$He.\cite{Taniguchi_2011} 
Due to aging, the surface area $S$ was reduced to 145~m$^{2}$, in the present 
measurements. 
Unlike the liquid measurements, the capillary is plumbed directly 
to the cell, since the thermal conductivity is not seriously high. 

The heat capacity was measured by a quasi-adiabatic  
heat-pulse technique up to 14~atoms/nm$^{2}$, which corresponds to 
$n_{\mathrm{C}}$\cite{Demura_2017}. 
The temperature of the sample cell was monitored with RuO$_{2}$, 
which were glued onto the bottom face of the cell. These were 
calibrated against a commercial calibrated RuO$_{2}$ thermometer. 
The thermal relaxation time from the cell to the stage was 360-5000~s, 
which was more than one order of magnitude larger than that in the 
cell, 40-240~s.

For the vapor pressure measurements, we adopted the cell which was used in the 
previous double torsional oscillator (DTO) measurements\cite{Taniguchi_2013}. 
As in the previous work\cite{Ikegami_2003}, the pressure was 
measured by means of the capacitive strain gauge with a membrane 
100$\mu$m in thickness whose one side was Au sputtered as an electrode. 
The pressure was calibrated against the $^{4}$He saturated vapor pressure. 
Its accuracy was 2$\times10^{-3}$~mbar. 
This gauge was attached directly to the cell, which was mounted on the 1K pot of 
the refrigerator, in order to avoid the temperature difference. 

\section{Results and Discussion}

\subsection{Isothermal compressibility and isosteric heat of sorption }
 \begin{figure}[t]
\includegraphics[width=16pc]{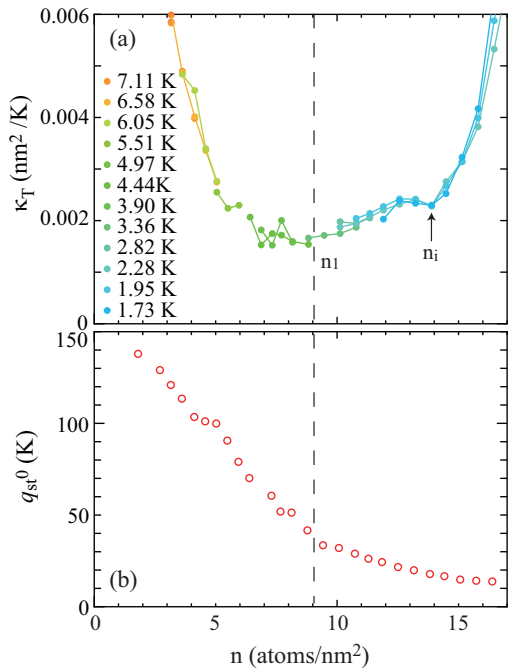}
\caption{\label{fig:epsart} (a) Two-dimensional isothermal compressibility $\kappa_{T}$ 
and (b) isosteric heat of sorption at absolute zero $q_{st}^{0}$ 
as a function of areal density. The vertical dashed line and the arrow show $n_{1}$ and $n_{i}$, 
respectively. 
}
\end{figure}

Vapor pressure ($P$) of the adsorbed $^{4}$He gives us the isothermal compressibility $\kappa_{T}$ and 
the isosteric heat of sorption $q_{st}$, which are useful to examine the change in film structure. 
Since $\kappa_{T}$ becomes small as the density is enhanced near the layer completion, 
its minimum is often used as a criterion of the completion of layers. 
$\kappa_{T}$ is deduced from the $P$- isotherm as 
\begin{equation}
\kappa_{T} = \frac{1}{n^2 k_{B} T}\left( \frac{\partial n}{\partial \ln P} \right),
\end{equation}
where $n$ is the areal density, $k_{B}$ is the Boltzmann constant, and $T$ is the temperature.

Figure 1 shows the obtained $\kappa_{T}$ as a function of areal density up to 16.4~atoms/nm$^{2}$, where the capillary condensation occurs. 
$\kappa_{T}$ shows a minimum at around 9$\pm 1$~atoms/nm$^{2}$, which we determine as the areal density 
of the first layer completion ($n_{1}$). Above $n_{1}$, $\kappa_{T}$ increases with increasing areal density, 
and turns to decrease at around 12.7$\pm0.3$~atoms/nm$^{2}$, and then has a small  
minimum at $n_{i}=$14~atoms/nm$^{2}$, which coincides with $n_{\mathrm{C}}$. 
Such a small minimum of $\kappa_{T}$ was also reported for $^{4}$He in 4.7- and 2.8-nm channels of FSM series by Ikegami et al.\cite{Ikegami_2003,Com0} Just above the areal density of the small minimum, the superfluid transition is commonly observed in both the present and the previous work. 
It indicates that the inert layer is slightly compressed as the areal density approaches $n_{\mathrm{C}}$.

On the other hand, from the temperature dependence of vapor pressure, we calculated $q_{st}$ as  
\begin{equation}
q_{st} = -k_B \frac{d \ln P}{d(1/T)}. 
\end{equation}
By subtracting the heat capacity of gas phase $^{4}$He, 
we estimate the isosteric heat of sorption at absolute zero, $q_{st}^{0} = q_{st}-(5/2)k_B T$, which 
corresponds to the depth of the adsorption potential\cite{Toda_2009}.  
The obtained $q_{st}^{0}$ is shown as a function of $n$ in Fig.~1(b). 
It is $\sim$140~K at 1.8~atoms/nm$^{2}$ and decreases monotonously with increasing areal 
density. It reaches $\sim$40~K at $n_{1}$, above around which its decrease becomes slow. 
The obtained value of $q_{st}^{0}$ is close to that of 4.7-nm channels of FSM, up to $n_{1}$\cite{Ikegami_2003}. 
It indicates that the difference in the channel size between 2.8 and 4.7~nm does not 
affect strongly on the adsorption potential in the submonolayer region.

\subsection{Heat capacity of inert layer $^{4}$He}

Figure~2 shows the heat capacity ($C$) for various areal densities as a function of 
$T$. Here, the heat capacity of the empty cell is subtracted. 
For 3.5~atoms/nm$^{2}$, a broad peak appears at around 1.1 K, in addition to the slope a little 
smaller than $T^{2}$. With increasing areal density, this peak shifts to the low-temperature side, with 
its height shrinking. Finally, it becomes unclear above 11.4~atoms/nm$^{2}$. Since the peak is very 
broad, we define $T_{\mathrm{p}}$ as the temperature where $C/T$ starts to 
deviate from the extrapolation of the low-temperature side, which gives us the lower limit of peak 
temperature. (see the inset of Fig.~2.) On the other hand, 
the temperature dependence at the high-temperature side keeps a little smaller than $T^{2}$ up to 
around 10.2~atoms/nm$^{2}$, and above that its slope slightly decreases. 

\begin{figure}[t]
\includegraphics[width=16pc]{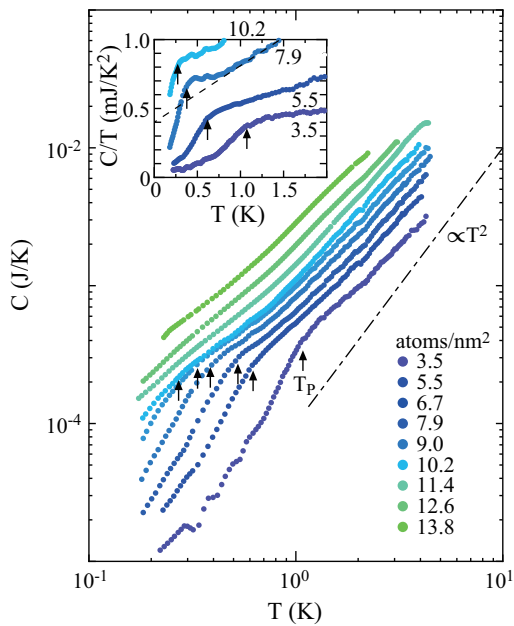}
\caption{\label{fig:1} Heat capacity of $^{4}$He as a function of temperature 
for various areal densities below 14~atoms/nm$^{2}$. The arrows point $T_{\mathrm{p}}$. 
The dot-dashed line is proportional to $T^{2}$. 
The figures are areal densities in the unit of atoms/nm$^{2}$. 
Inset: $C/T$ as a function of $T$ below 9.0~atoms/nm$^{2}$. 
The dotted line is the extrapolation from the high temperature side of the peak.
}
\end{figure}

As is clear from the inset of Fig.~2, the heat capacity at the temperature fully higher than  
$T_{\mathrm{p}}$ can be explained by the sum of $T$-linear ($y$-intercept) and $T$-squared (slope) 
terms. In addition, there is a peak at around $T_{\mathrm{p}}$. 
It should be noted that the $y$-intercept of the extrapolation from the heat capacity near the 
lowest temperature is lower than that of the extrapolation from the high-temperature side of 
$T_{\mathrm{p}}$ (dashed line in the inset of Fig.~2.). There is a possibility that the difference of 
$T$-linear term comes from the contribution of the excited atoms at around $T_{\mathrm{p}}$. 

Therefore, we set the heat capacity model as 
\begin{eqnarray}
C &=& (A_{l}T + B T^{2})  + A_{\mathrm {exc}} T +  D \frac{(\beta \Delta E)^2}{(\cosh(\beta \Delta E))^{2}}, \\
A_{\mathrm {exc}} &=& A_{\mathrm {exc0}} \frac{\exp (-\beta \Delta E)}{\exp (\beta \Delta E) + \exp (-\beta \Delta E)}, 
\end{eqnarray}
where $\beta = 1/k_BT$. 
Here, the first parenthesis, the second and the third terms correspond to the heat capacity of the 
localized solid, the excited fluid $^{4}$He, and a Schottky peak due 
to the excitation, respectively. 
Regarding their origin, Tait and Reppy suggested that $A_{l}$ and $B$ come from the amorphous property 
and 2D phonon of the localized solid, and that $A_{\mathrm{exc}}$, from 2D Bose gas. 
$A_{\mathrm{exc0}}$ corresponds to the slope when $^{4}$He atoms are fully excited. 
$2\Delta E$ is the energy gap between the localized solid and the 2D gas states.   
Based on this model, we discuss the film structure later.

Figure~3 shows the fitted curves of 5.5~atoms/nm$^{2}$ as an example. 
The heat capacity is well reproduced over the entire temperature range. 
At around the lowest temperature, $T$-linear term becomes dominant, while near the highest 
temperature $T^{2}$-term is dominant. At around $T_{\mathrm{p}}$, the contribution of 
Schottky term becomes large, and the slope of $T$-linear term increases due to the contribution 
of $A_{\mathrm{exc}}T$. 
The heat capacity for all areal densities between 3.5 and 13.8~atoms/nm$^{2}$ are fitted well to 
the eq.~(3).

\begin{figure}[t]
\includegraphics[width=16pc]{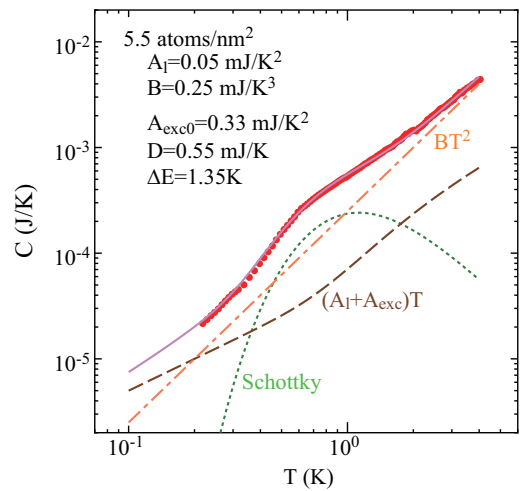}
\caption{\label{fig:2} Heat capacity of $^{4}$He at 5.5~atoms/nm$^{2}$ as a 
function of temperature. The solid curve is the fitted curve to eq. (3). 
The dashed, dot-dashed, and dotted curves correspond to the $T$-linear, 
$T$-squared and the Schottky terms in the Eq.~3, respectively. }
\end{figure}

\subsection{Property of the localized solid}

It is well known that the amorphous property of the localized solid is characterized by the $T$-linear 
heat capacity due to the quantum tunneling between different states in the two-level systems (TLS). 
Figure~4(a) shows the fitting results of $A_{l}$ as a function of areal density. It is quite small up to around 
9~atoms/nm$^{2}$ and then soars with increasing areal density. 

The coefficient of $T$-linear term for amorphous solid is described as $\pi^2/6\cdot D_{0}k_B$, 
where $D_{0}$ is the density of states. 
When the TLS is generated by the distribution of adsorption potential, $D_{0}$ is often approximated 
as \cite{Tait_1979,Toda_2009}
\begin{equation}
D_{0} = \left(\frac{\Delta q_{st}^{0}}{\Delta n} \right)^{-1}. 
\end{equation}
Using $q^{0}_{st}$ in Fig.~1(b), we estimated the coefficient of $T$-linear heat capacity, $A_{qst}$, 
which is shown in Fig.~4(a) for comparison. The calculated $A_{qst}$ shows the same 
areal density dependence as that of $A_{l}$, except that it is slightly larger than $A_{l}$. 
The semiquantitative agreement means that $A_{l}T$ term is well explained by the contribution of 
amorphous solid.

\begin{figure}[t]
\includegraphics[width=15pc]{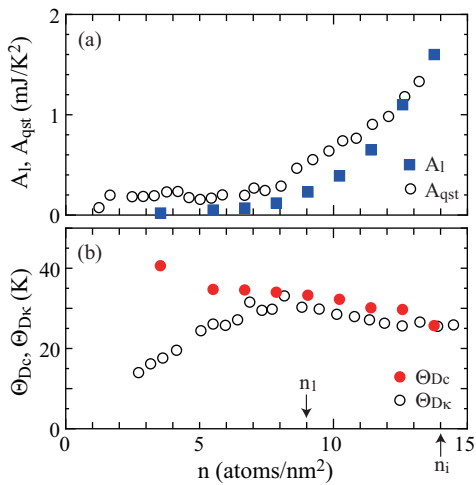}
\caption{\label{fig:epsart} (a) Fitting results of $A_{\mathrm{l}}$ and the calculated $A_{qst}$ 
as a function of areal density. 
(b) Debye temperatures obtained from fitting parameter $B$($\Theta_{Dc}$) and 
from $\kappa_{T}$ ($\Theta_{D\kappa}$) as a function of areal density. 
}
\end{figure}

Next, we consider the $T^{2}$ term. 
From the fitting results of $B$, we calculate the Debye temperature $\Theta_{\mathrm{Dc}}$ as
$\Theta_{\mathrm{Dc}}=(28.8 Nk_B/B)^{0.5}$,
where $N$ is the number of solid $^{4}$He. Here, we approximate $N$ by $n$ multiplied by the surface 
area $S$, neglecting the number of excited $^{4}$He. 
The areal density dependence of $\Theta_{\mathrm{Dc}}$ is shown in Fig.~4(b). 
$\Theta_{\mathrm{Dc}}$ is 43~K for 3.5~atoms/nm$^{2}$, and drops to $\sim35$~K at 5.5~atoms/nm$^{2}$
and above that decreases slightly with increasing areal density. The value at around $n_{1}$ (34~K) is close 
to those for other substrates 
such as Cu (32~K)\cite{Roy_1971} and graphite (33.0~K at 9.67~atoms/nm$^{2}$)\cite{Bretz_1973}.

The Debye temperature can be also deduced from the phonon velocity $v_{\mathrm{p}}$ as 
$\Theta_{\mathrm{D}\kappa}= (hv_{\mathrm{p}}/k_B) (n/\pi)^{0.5}$. 
The phonon velocity is related to 
the adiabatic compressibility $\kappa_{S}$ as 
$v_{\mathrm{p}}= \sqrt{1/(mn\kappa_{S})}$. 
Here, we evaluate $\Theta_{\mathrm{D}\kappa}$ by assuming that 
$\kappa_{T}$ approximately equals $\kappa_{S}$.\cite{Toda_2009}
$\Theta_{\mathrm{D}\kappa}$ increases with increasing areal density and then turns to decrease at 
$\sim$8~atoms/nm$^{2}$, above which $\Theta_{\mathrm{D}\kappa}$ agrees well with $\Theta_{\mathrm{Dc}}$. 
The reason why $\Theta_{\mathrm{D}\kappa}$ is smaller than $\Theta_{\mathrm{Dc}}$ 
below 8~atoms/nm$^{2}$ is left as a question. However, one possibility is that in this 
areal density region, the solid islands are considered to be independent of each other. 
In this situation, $\kappa_T$ may not reflect the compressibility of the solid $^{4}$He 
itself.


The monotonous decrease of $\Theta_{\mathrm{Dc}}$ with increasing areal density is not intuitive. 
Naively, when the number of atoms increases, the film is expected to be compressed, making the film 
stiff, i.e. raising the Debye temperature. In fact, for monolayer $^{4}$He adsorbed on graphite, 
the increase of the Debye temprature is reported\cite{Bretz_1973}. 
To understand the $\Theta_{\mathrm{Dc}}$ behavior, 
it is necessary to consider the distribution of adsorption potential, which generates the lateral pressure to compress the film. 
The calculation by Rossi et al. shows that the adsorption potential depends on the azimuthal direction in 
the cross section, and that its distribution decreases with increasing the film thickness i.e. the areal density. 
It means that with increasing areal density, the lateral pressure lowers, leading to the decrease of 
$\Theta_{\mathrm{Dc}}$. 

\subsection{Excitation to the 2D gas state}

First, we focus on the Schottky peak term. 
In Fig.~5(a) and (b), the fitting parameters $\Delta E$, and $D$ divided by $k_B S$ are plotted 
as a function of $n$, respectively, up to 11.4~atoms/nm$^{2}$, above which the peak cannot be 
identified. Here, $D/(k_BS)$ corresponds to the number of atoms that 
can be excited in the unit of areal density. $\Delta E$ is 2.2~K for 3.5~atoms/nm$^{2}$, and 
decreases with increasing areal density, at a decreasing rate.  
The areal density dependence and the order of magnitude of $\Delta E$ agree with 
those for $^{4}$He on vycor glass, supporting that the 
adsorbed $^{4}$He on the present substrate are also excited to 
the gas state. 

On the other hand, $D/(k_BS)$ is at most $\sim$10~\% for 3.5~atoms/nm$^{2}$, and 
decreases with increasing areal density. Above $n_{1}$, it decreases at an accelerated 
rate and seems to go to zero at around 13~atoms/nm$^{2}$. 
The fitting results indicate that 2D gas disappears just before the inert layer completes,  contrary to the conventional understanding that the 2D gas phase below 
$n_{\mathrm{C}}$ is continuous to the normal phase above the superfluid transition 
temperature above $n_{\mathrm{C}}$.  

\begin{figure}[t]
\includegraphics[width=18pc]{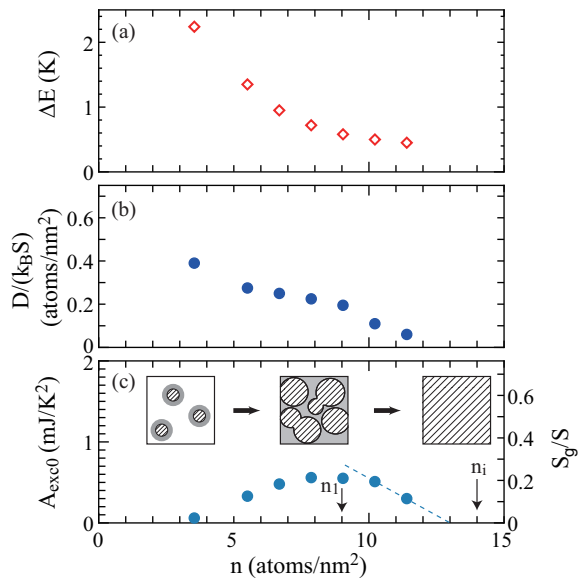}
\caption{\label{fig:epsart} Fitting results of (a) $\Delta E$ (b) $D/(k_BS)$, 
(c) $A_{\mathrm{exc0}}$ are shown as a function of areal density. 
In (c), the dotted line is the extrapolation of the decrease above 10~atoms/nm$^{2}$. 
The right vertical axis of (c) is $S_{\mathrm{g}}/S$. The inset pictures of (c) show the 
schematic top views of the film at around 3.5~atoms/nm$^{2}$, $n_{1}$, and $n_{i}$ from left to right. 
The hatching and gray areas corresponds to the solid and the fluid areas, respectively. }
\end{figure}

To examine the relation between the amount of 2D gas $^{4}$He and the magnitude of 
$T$-linear heat capacity, 
we plot $A_{\mathrm{exc0}}$ against $n$ in Fig.~5(c). 
It is 0.06~mJ/K$^{2}$ for 3.5~atoms/nm$^{2}$, and 
increases with increasing areal density at first. Then, it becomes almost constant between 8 and 10~atoms/nm$^{2}$, and turns to decrease. 
The important thing is that $A_{\mathrm{exc0}}$ is not proportional to $D/k_BS$, i.e. 
the amount of 2D gas at a fully high temperature. 

In the case of  an ideal 2D Bose gas, the heat capacity is $T$-linear and depends 
not on the number of atoms but on the surface area that the gas covers ($S_{g}$). 
The coefficient is described as $ (mS_{g})/(2\pi \hbar ^2) \cdot k_{B}^{2}$, where $m$ is the mass of $^{4}$He atom\cite{Daunt_1972}. When the entire surface area is covered, 
it becomes 2.6~mJ/K$^{2}$, which is much larger than the obtained 
$A_{\mathrm{exc}}$, suggesting that only a part of surface is occupied by the excited gas. 

From the viewpoint of change in $S_{g}/S$, we consider the structure of the inert layer. 
Here, we define $S_{g}/S$ as $A_{\mathrm{exc0}}/(2.6$ mJ/K$^{2})$, which is shown as the right vertical 
axis of Fig.~5(c). On the other hand, the surface area covered by the localized solid ($S_{s}$) is approximated 
by $(n/n_{1})S$ in the submonolayer region, 
while above $n_{1}$ it is defined as $S_{s}=S-S_{g}$. 
$S_{g}$ increases in conjunction with $n$, i.e., $S_{s}$, at first, and then its increase stops at around 
8~atoms/nm$^{2}$, where $S_{g}+S_{s}$ reaches $S$. The almost constant $S_{g}$ between 8 and 
10~atoms/nm$^{2}$ indicates that the increase of $n$ causes promotion of the gas 
$^{4}$He to the overlayer. It is thought to be accompanied by the compression of the first layer, since 
$\kappa_{T}$ takes minimum in the same areal density region. Above that, $S_{g}/S$ turns to decrease and 
is likely to go to zero at around 13~atoms/nm$^{2}$. It is understood as the increase of density in the overlayer 
raises the ratio of $S_{s}$ to $S_{g}$ and finally lets the entire overlayer solidified. 
It is consistent with the decrease of $\kappa_{T}$ above around 13~atoms/nm$^{2}$, indicating that the compression starts at the end of the coexistence of the gas and solid.  



As the origin which limits $S_{g}$, the lateral pressure due to the distribution of long-range 
adsorption potential is suggested\cite{Tait_1979}. In the case of vycor glass, it is thought 
that the distribution of pore size generates the distribution of the adsorption potential. 
Although the channel size of FSM is uniform, 
it is reported that the adsorption potential of a hexagonal channel has an azimuthal dependence in 
cross-section\cite{Rossi_2006}. Its amplitude is several tens K near the pore wall, and 
decreases as the distance from the wall increases, and becomes zero when the 
film thickness reaches 0.75~atoms/nm$^{2}$, at which superfluid appears\cite{Demura_2017}.  
The amplitude in the submonolayer region is larger than the obtained $\Delta_E$, 
letting the excited gas locate only around the rim of the solid islands. 
It means that the excited gas does not spread uniformly but is condensed. 
Therefore, we call it fluid in the following discussion.


\subsection{Phase diagram}

\begin{figure}[t]
\includegraphics[width=15pc]{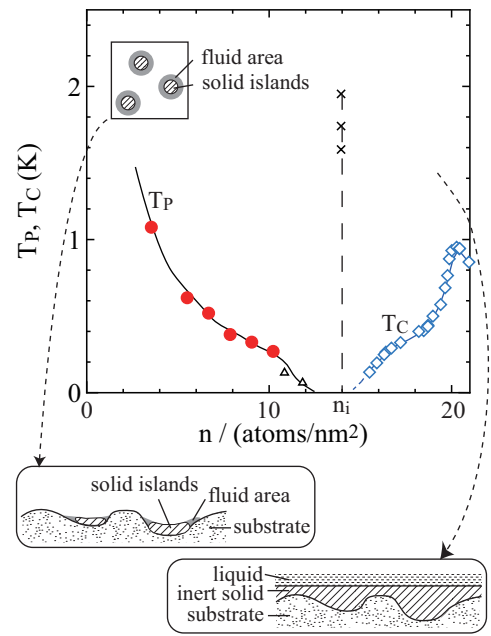}
\caption{\label{fig:epsart} $T_{\mathrm{P}}$ and the temperature of dissipation peak due to 
the superfluid in the channel ($T_{\mathrm{C}}$) as a function of areal density. ($\triangle$) is the 
data of $T_{\mathrm{P}}$ from Ref.~22. The temperatures at 
which the second minimum in $\kappa_{T}$ is observed are shown as ($\times$). Inset: the schematic top 
view of the film above around $T_{\mathrm{P}}$. The two lower left and right panels are cross-section images of the films above around $T_{\mathrm{P}}$ and above $T_{\mathrm{C}}$, respectively. }
\end{figure}

We discuss the phase diagram of $^{4}$He film, based on the foregoing discussion. 
Figure~6 summarizes the typical temperatures of phase separations, $T_{\mathrm{P}}$ and 
$T_{\mathrm{C}}$ as a function of areal density. 
In the low areal density region, $^{4}$He atoms are adsorbed on the areas with a deep adsorption 
potential and forms solid islands at fully low temperature. At around $T_{\mathrm{P}}$, a small part 
of the localized $^{4}$He is excited and forms a fluid area surrounding the solid islands. 
With increasing areal density, $T_{\mathrm{P}}$ decreases due to the decrease of adsorption potential 
distribution, and finally disappears at around 13~atoms/nm$^{2}$. Above this areal density, 
instead of the decrease in fluid areas, the compression of the inert layer composed of only solid occurs. 
Then above $n_{i}$, the liquid layer appears on top of the inert layer, which shows superfluidity at low 
temperature. 

This phase diagram is similar to the one probably characteristic of adsorbed $^{4}$He 
on heterogeneous substrates proposed by Tait and Reppy. 
The only thing that is different is that the fluid and the liquid phases are not continuous. 
We consider that after the inert layer completion at which the lateral pressure becomes fully small, 
the uniform 2D liquid phase appears. This idea is consistent with the fact that the superfluid fraction 
is almost proportional to $(n-n_{i})$.  


\section{Summary}

In summary, we have studied the structure of inert layer $^{4}$He in a 2.8-nm channel of FSM-16, 
based on the vapor pressure and heat capacity data. 
We analyzed the heat capacity of the inert layer based on the model including the contribution of the 
localized solid and excited fluid. 
The results support the picture that the excited fluid coexists with the localized solid at high temperature. 
With increasing areal density, the amount of excited fluid is decreased and likely to go to zero 
just below $n_{\mathrm{C}}$(=$n_{i}$). 
It suggests the possibility that the normal fluid phase below $n_{\mathrm{C}}$ is not continuous to 
the liquid phase above $n_{\mathrm{C}}$. 

\begin{acknowledgments}
The work was partly supported by JSPS KAKENHI Grant No. 26400352 and 
No. 18K03535. We thank S. Inagaki for the supply of the FSM-16. 
\end{acknowledgments}

\nocite{*}


\end{document}